\documentclass[12pt,a4paper]{article}

\usepackage[T1]{fontenc}
\usepackage[utf8]{inputenc}
\usepackage{lmodern}
\usepackage[margin=2.5cm]{geometry}
\usepackage{microtype}
\usepackage{setspace}
\usepackage{amsmath,amssymb}

\usepackage{graphicx}
\usepackage{tikz}
\usetikzlibrary{arrows.meta, shapes.geometric, positioning, fit, backgrounds, calc}
\usepackage{subcaption}
\usepackage{float}
\usepackage{booktabs}
\usepackage{array}
\usepackage{multirow}

\usepackage[usenames,dvipsnames,svgnames,x11names]{xcolor}
\definecolor{darkblue}{HTML}{2166AC}
\definecolor{medblue}{HTML}{74ADD1}
\definecolor{myorange}{HTML}{F46D43}
\definecolor{mygrey}{HTML}{AAAAAA}
\definecolor{myteal}{HTML}{009688}

\usepackage[colorlinks=true,linkcolor=darkblue,citecolor=darkblue,urlcolor=darkblue]{hyperref}

\usepackage[numbers,sort&compress]{natbib}

\usepackage{listings}
\newcommand{\HIGH}{\textbf{HIGH}}
\newcommand{\MED}{\textbf{MEDIUM}}
\newcommand{\LOW}{\textbf{LOW}}
\newcommand{\sanadlinks}{\texttt{sanad\_links}}
\newcommand{\narratorlinks}{\texttt{narrator\_links}}

\begin{document}

\title{\textbf{Linking Hadith Narrator Identities Across
       Heterogeneous Arabic Biographical Databases: A Multi-Signal Entity Resolution Pipeline}}

\author{Taufiq Wirahman\\
        \small Research Center for Computing, National Research and Innovation Agency (BRIN),\\ \small Soekarno Science and Technology Zone, Cibinong, West Java, Indonesia\\[2pt] \small Department of Mathematics, Faculty of Sciences,\\ \small Universiti Teknologi Malaysia, Johor Bahru, Malaysia\\[2pt] \small \href{mailto:taufiq.wirahman@brin.go.id}{taufiq.wirahman@brin.go.id}}

\date{June 2026}
\maketitle

% ============================================================
\begin{abstract}
The transmission chains (\textit{sanad}) of Islamic Hadith literature encode relationships among tens of thousands of historical narrators whose biographical records are dispersed across independently maintained digital databases that share no common identifier. We present a two-phase entity resolution pipeline that links narrator names from the Sanadset 650K corpus---650,986 Hadith records from 926 books containing 185,216 unique narrator name variants---to two biographical databases: Hadithtransmitters (hawramani; 100,915 entries) and Muslimscholars (25,247 entries). Phase~1 matches Sanadset names to hawramani using name-only similarity (Sanadset carries no metadata), yielding 94,628 links (51.1\%; \HIGH{} 39,938 / \MED{} 54,690). Phase~2 cross-references hawramani against muslimscholars via a weighted multi-signal function combining name similarity, death-year proximity, and reliability grade polarity, yielding 95,573 links (94.7\% of hawramani; \HIGH{} 18,245 / \MED{} 71,546 / \LOW{} 5,782). Chaining the two phases gives Sanadset narrators transitive access to muslimscholars data. The linked data enable construction of a 185,216-node, 814,093-edge directed transmission graph enriched with cross-source biographical metadata. The annotated link corpora and enriched graph are released as open resources.
\end{abstract}


% ============================================================
\section{Introduction}
\label{sec:intro}

Islamic Hadith scholarship rests on a discipline known as \textit{`ilm al-rij\={a}l}---the science of narrator evaluation. For over a millennium, scholars have manually cross-referenced narrator biographies recorded in Arabic \textit{tabaqa\={t}} (biographical dictionaries) to assess the reliability of Hadith transmission chains (\textit{sanad}). A single narrator may appear under different name forms across sources: a kunyah (\textit{Ab\={u} \d{H}an\={\i}fa}) in one database, a full nasab chain (\textit{al-Nu\`m\={a}n ibn Th\={a}bit ibn Z\={u}t\={a}}) in another, with or without diacritical marks, and in different grammatical cases depending on syntactic position within the chain.

The digitization of Hadith literature has produced several large, independently maintained databases. Sanadset 650K \citep{mghari2022sanadset} provides 650,986 structured Hadith records from 926 Arabic books with XML-tagged narrator chains, yielding 185,216 unique narrator name variants. The Hadithtransmitters platform (hawramani) hosts biographical records for 100,915 narrators including death years, reliability grades, and teacher--student links. Muslimscholars.info maintains structured profiles for 25,247 Islamic scholars. Each database uses different name representations, coverage criteria, and data models, and none is linked to the others at the entity level. This isolation imposes a structural ceiling on computational Hadith research: social network analyses of transmission chains \citep{saeed2021sna,alam2021sna} reveal rich graph-theoretic properties but are constrained to single collections with no biographical metadata on narrator nodes; narrator disambiguation systems \citep{mahmoud2022arsanad,mahmoud2024arsanadv2} achieve high classification accuracy but operate entirely within a single biographical authority.

Existing approaches to Arabic entity resolution target modern Wikipedia-aligned knowledge bases and cannot address the distinctive challenges of historical Islamic personal names: multi-token nasab chains of variable length, kunyah/ism alternation, medieval spelling conventions, and morphological case variation \citep{elzanfaly2016arabic,mohasseb2013hybrid}. While probabilistic multi-signal record linkage provides a theoretical foundation \citep{christen2012data}, no domain-adapted pipeline existed for cross-source Hadith narrator linking prior to this work---a gap consistently identified in prior surveys \citep{mhamedi2023computational}. Section~\ref{sec:related} reviews the relevant literature in detail.

This paper addresses this gap with three contributions: (i) a two-phase entity resolution pipeline combining a name-only phase (Sanadset~$\to$~hawramani) and a multi-signal phase (hawramani~$\leftrightarrow$~muslimscholars) using name similarity, death-year proximity, and grade polarity with domain-specific Arabic normalization; (ii) two annotated link corpora---\sanadlinks{} (Sanadset-to-hawramani) and \narratorlinks{} (hawramani-to-muslimscholars)---both stratified into confidence tiers; and (iii) a large-scale directed transmission graph enriched with cross-source biographical metadata.

This paper is organized as follows. Section~\ref{sec:related} reviews related work; Sections~\ref{sec:data}--\ref{sec:results} describe the databases, pipeline, and results; Section~\ref{sec:discussion} discusses limitations and future directions; Section~\ref{sec:conclusion} concludes.

% ============================================================
\section{Related Work}
\label{sec:related}

\subsection{Computational Hadith and Digital Islamic Humanities}

Sanadset 650K \citep{mghari2022sanadset} provides 650,986 Hadith records from 926 books with XML markup tagging narrator mentions within chain wrappers, and forms the source corpus for the entity resolution task in this paper. Narrator disambiguation has been formalized as multi-class classification: AR-Sanad 280K \citep{mahmoud2022arsanad} introduced 279,625 synthetic sanad sequences with 18,298 narrator classes and established AraBERT \citep{antoun2020arabert} as the dominant baseline model. AR-Sanad v2 \citep{mahmoud2024arsanadv2} extended this by integrating a NarratorsKG into the disambiguation pipeline. Both treat disambiguation as a \emph{closed-world} problem requiring the narrator to exist within a predefined class set. Our work addresses the complementary \emph{open-world} problem: given a narrator name from Sanadset, find its matching record in external databases that use different name representations.

Narrator2Vec \citep{mahmoud2023narrator2vec} represents narrator names as continuous vectors trained on the Sanadset corpus, providing a soft name- similarity signal without explicit entity links or confidence scores. Social network analyses \citep{saeed2021sna,alam2021sna} reveal scale-free degree distributions in Hadith transmission chains and a geographic shift in narrator hubs over generations. Multi-IsnadSet \citep{multiisnadset2024} released a directed graph from Sahih Muslim with 2,092 narrator nodes and 77,797 edges; the graph constructed in this paper spans 926 books and 185,216 nodes. Knowledge graph approaches \citep{kamran2023semantic, kamran2026enrichment} model Hadith structure as RDF triples; they do not address cross-source narrator entity resolution. \citet{romanov2017algorithmic} and \citet{romanov2022topological} established computational methods for scholarly networks from Arabic biographical collections; \citet{mhamedi2023computational} surveyed sanad methodologies and identified cross-source narrator identity resolution as an open problem.

\subsection{Arabic Entity Resolution}

Arabic named entity disambiguation using Wikipedia-aligned knowledge bases \citep{elzanfaly2016arabic} relies on entity descriptions unavailable for historical Islamic narrators. Hybrid Arabic name matching \citep{mohasseb2013hybrid} combines token-level Jaccard similarity with edit distance after normalization of Alef variants and diacritics---techniques directly relevant to Section~\ref{sec:norm}. The probabilistic record linkage framework of \citet{christen2012data} provides the theoretical grounding for the weighted multi-signal scoring in Section~\ref{sec:phase2}. Arabic knowledge graph construction faces challenges compounded in the historical Hadith domain: sparse Arabic coverage in online encyclopedias, morphological richness, and Alef-variant proliferation \citep{elzanfaly2017arabic}. Applying AraBERT \citep{antoun2020arabert} to cross-source linking would require fine-tuned alignment pairs that do not exist prior to linking---the bootstrapping problem our pipeline addresses by producing the first such corpus.

% ============================================================
\section{Data Sources}
\label{sec:data}

\subsection{Sanadset 650K}

Sanadset 650K \citep{mghari2022sanadset} is a structured corpus of 650,986 Hadith records from 926 historical Arabic books. Each record carries the full Hadith text with XML-like markup tagging narrator chains (\texttt{<SANAD>}) and individual narrator mentions (\texttt{<NAR>}). After exploding narrator lists and normalizing names (Section~\ref{sec:norm}), the corpus yields \textbf{185,216} unique bare narrator name variants, participating in approximately 4.8 million narrator--Hadith occurrence pairs. Sanadset carries no biographical metadata (death years, grades) for narrator names---a constraint that determines the design of Phase~1.

\subsection{Hadithtransmitters (Hawramani)}

The Hadithtransmitters platform~\citep{hawramani2026} hosts biographical records for \textbf{100,915} Hadith narrators compiled from classical Islamic biographical sources. Each record provides: full Arabic name, URL-slug identifier, estimated death year (present for a subset of narrators; see Section~\ref{sec:scoring-death}), reliability grades from multiple classical \textit{rij\={a}l} scholars, source citation count, and hyperlinked teacher and student name lists. The database was scraped via breadth-first search (BFS) seeded from known narrator URL patterns, with three HTTP requests per narrator (main biography, teacher list, student list). Teacher and student IDs encountered on each page were added to the BFS queue, enabling organic discovery of all narrators without exhaustive range scanning. The complete crawl covered 100,915 records.

\subsection{Muslimscholars}

Muslimscholars.info~\citep{muslimscholars2026} hosts profiles for \textbf{25,247} Islamic scholars. Each record provides Arabic and English name forms, estimated death year (present for 21.7\% of muslimscholars entries appearing in Phase~2 matched pairs; see Table~\ref{tab:p2}), school of jurisprudence (\textit{mazhab}), and teacher/student links as integer IDs. Compared to hawramani, muslimscholars covers a narrower but well-documented subset, and provides English romanizations useful as a secondary signal. Scraping used the same BFS strategy, seeded from compact known ID ranges with subsequent queue expansion via linked IDs, reaching high-ID records (up to ID~151,001) without scanning the estimated 110,000 empty slots between known ranges.

\subsection{Narrators Knowledge Base (narrators\_db)}

The narrators\_db from AR-Sanad 280K \citep{mahmoud2022arsanad} provides structured biographical metadata for \textbf{18,298} narrators: Ibn Hajar reliability rank (12 classes), \textit{shuhra} (epithet), \textit{laqab} (honorific), nasab chain, family ties, and school of jurisprudence. Death year fields are present for all 18,298 narrators as free-form Arabic text strings (e.g., a year numeral followed by the Arabic Hijri abbreviation, or a date-range such as ``AH~141--150''); 8,802 carry a parseable Hijri year in the range AH~1--500. This resource provides death-year data for Figure~\ref{fig:timeline}.

% ============================================================
\section{Methodology}
\label{sec:method}

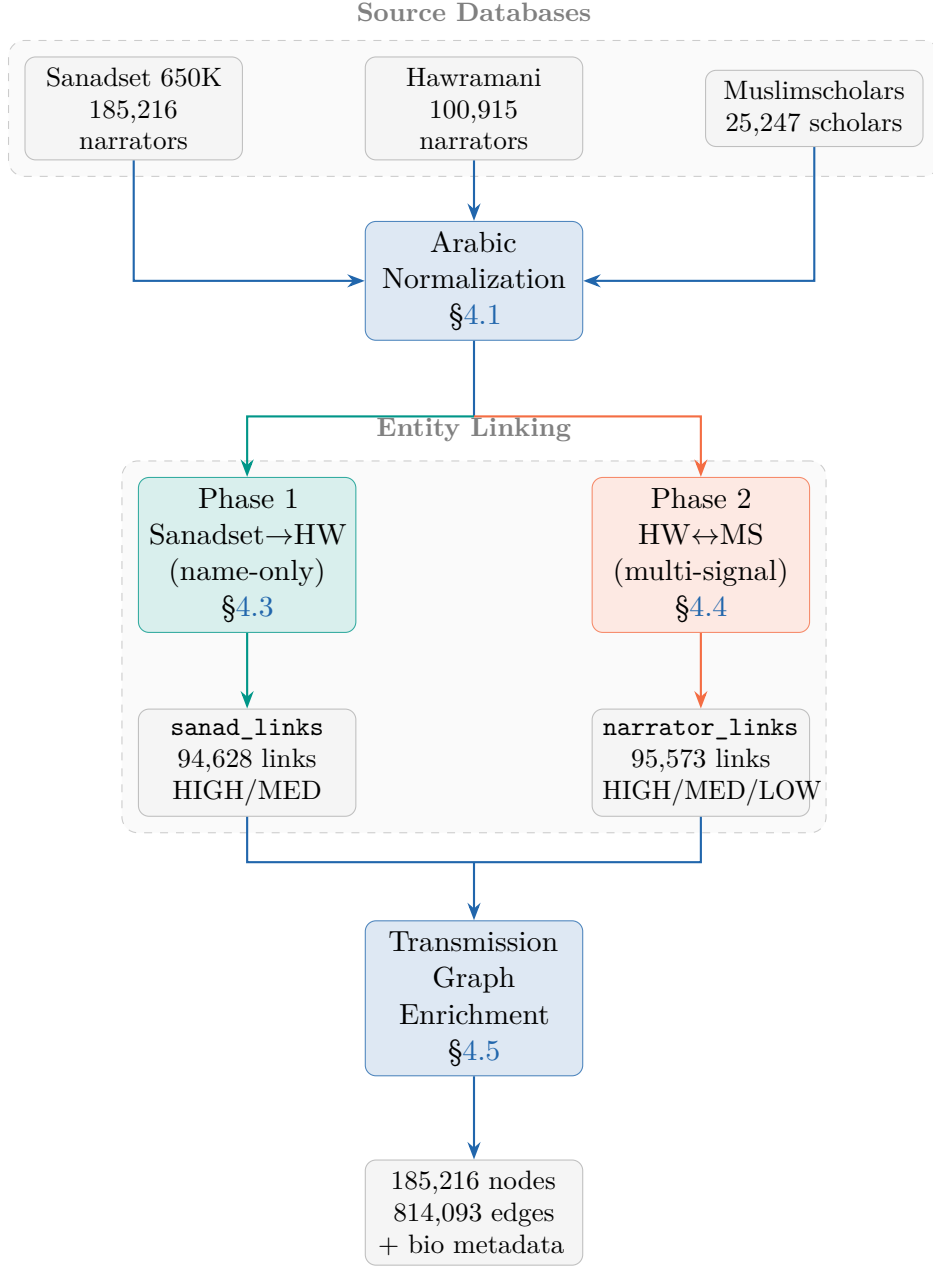
\begin{figure}[!ht]
\centering
% ── TikZ pipeline diagram ──────────────────────────────────────────────────────
\begin{tikzpicture}[
  font=\small, box/.style  = {draw, rounded corners=4pt, fill=#1!15, draw=#1!80, minimum width=2.6cm, minimum height=0.85cm, align=center, text width=2.6cm}, dbox/.style = {draw, rounded corners=4pt, fill=gray!8, draw=gray!50, minimum width=2.6cm, minimum height=0.75cm, align=center, text width=2.6cm, font=\footnotesize}, arr/.style  = {-Stealth, thick, #1}, grp/.style  = {draw=gray!40, dashed, rounded corners=6pt, inner sep=6pt, fill=gray!4}, lbl/.style  = {font=\bfseries\footnotesize, #1}, ]

% ── Nodes ──────────────────────────────────────────────────────────────────────
% Row 0: source databases (absolute anchors)
\node[dbox] (sanad) at (0,0)    {Sanadset 650K\\185,216 narrators};
\node[dbox] (hw)    at (4.5,0)  {Hawramani\\100,915 narrators};
\node[dbox] (ms)    at (9,0)    {Muslimscholars\\25,247 scholars};

% Row 1: shared normalization — 0.8 cm gap below hw.south
\node[box=darkblue, below=0.8cm of hw] (norm)
  {Arabic\\Normalization\\\S\ref{sec:norm}};

% Row 2: linking phases — 1.8 cm gap below norm.south, offset ±3 cm in x
\node[box=myteal,   below=1.8cm of norm, xshift=-3cm] (p1)
  {Phase 1\\Sanadset$\to$HW\\(name-only)\\\S\ref{sec:phase1}};
\node[box=myorange, below=1.8cm of norm, xshift= 3cm] (p2)
  {Phase 2\\HW$\leftrightarrow$MS\\(multi-signal)\\\S\ref{sec:phase2}};

% Row 3: link outputs — 1.0 cm gap below p1/p2.south
\node[dbox, below=1.0cm of p1] (sl)
  {\sanadlinks\\94,628 links\\HIGH/MED};
\node[dbox, below=1.0cm of p2] (nl)
  {\narratorlinks\\95,573 links\\HIGH/MED/LOW};

% merge coordinate first (needed for graph placement)
\coordinate (merge) at ($(sl.south)+(3.0,-0.6)$);

% Row 4: graph — 1.8 cm below merge (gives clear arrow stem merge→graph)
\node[box=darkblue] (graph) at ($(merge)+(0,-1.8)$)
  {Transmission Graph\\Enrichment\\\S\ref{sec:graph}};

% Row 5: enriched output — 1.1 cm gap below graph.south
\node[dbox, below=1.1cm of graph] (gout)
  {185,216 nodes\\814,093 edges\\+ bio metadata};

% ── Arrows ─────────────────────────────────────────────────────────────────────
% Sources → Normalization
\draw[arr=darkblue] (sanad.south) |- (norm.west);
\draw[arr=darkblue] (hw.south)    -- (norm.north);
\draw[arr=darkblue] (ms.south)    |- (norm.east);

% Normalization → Phases via fork (1.0 cm below norm.south)
\coordinate (fork) at ($(norm.south)+(0,-1.0)$);
\draw[darkblue, thick] (norm.south) -- (fork);
\draw[arr=myteal]   (fork) -| (p1.north);
\draw[arr=myorange] (fork) -| (p2.north);

% Phases → Link outputs
\draw[arr=myteal]   (p1.south) -- (sl.north);
\draw[arr=myorange] (p2.south) -- (nl.north);

% Link outputs → Graph via merge
\draw[darkblue, thick] (sl.south) |- (merge);
\draw[darkblue, thick] (nl.south) |- (merge);
\draw[arr=darkblue] (merge) -- (graph.north);

% Graph → enriched output
\draw[arr=darkblue] (graph.south) -- (gout.north);

% ── Background groups ──────────────────────────────────────────────────────────
\begin{scope}[on background layer]
  \node[grp, fit=(sanad)(hw)(ms)]  (srcgrp)   {};
  \node[grp, fit=(p1)(p2)(sl)(nl)] (matchgrp) {};
\end{scope}
\node[above=3pt of srcgrp,   lbl=gray] {Source Databases};
\node[above=3pt of matchgrp, lbl=gray] {Entity Linking};

\end{tikzpicture}
\caption{The two-phase narrator entity resolution and graph enrichment pipeline.}
\label{fig:pipeline}
\end{figure}

Figure~\ref{fig:pipeline} illustrates the complete pipeline. Arabic normalization is applied to all three databases before any comparison (Section~\ref{sec:norm}). Phase~1 links Sanadset names to hawramani using name-only matching (Section~\ref{sec:phase1}); Phase~2 cross-references hawramani against muslimscholars using a multi-signal weighted score (Section~\ref{sec:phase2}). Both link corpora feed the transmission graph (Section~\ref{sec:graph}), where Phase~1 and Phase~2 links are transitively composed to give Sanadset narrators access to muslimscholars data.

% ── 4.1 Normalization ─────────────────────────────────────────────────────────
\subsection{Arabic Name Normalization}
\label{sec:norm}

Historical Arabic personal names require normalization beyond standard Arabic NLP preprocessing. The shared normalization procedure $\mathrm{norm}(t)$ applies four transformations in order:

\begin{enumerate}
  \item \textbf{Diacritic removal.} Arabic tashkeel characters
        (U+064B--U+065F and scattered supplementary codepoints) are stripped by targeting only the diacritic Unicode ranges. A non-trivial implementation hazard arises here: a naive codepoint range spanning U+0610--U+064A encompasses all Arabic consonants in addition to diacritics. An earlier implementation used this overly broad range and stripped all Arabic letters from input strings, causing every record to match every other record. The corrected procedure targets diacritics only via explicit, disjoint codepoint ranges.

  \item \textbf{Alef unification.} The five Alef variants---Alef (U+0627),
        Alef with hamza above (U+0623), Alef with hamza below (U+0625), Alef with madda (U+0622), and Alef wasla (U+0671)---are all mapped to bare Alef (U+0627).

  \item \textbf{Ta marbuta and Alef maqsura.} Ta marbuta
        (ta marb\={u}ta, U+0629) is normalized to ha (U+0647); Alef maqsura (alef maqsura, U+0649) to ya (U+064A).

  \item \textbf{Tatweel and whitespace.} Tatweel (U+0640) is stripped;
        multiple whitespace characters are collapsed to single spaces.
\end{enumerate}

The combined procedure reduces false negatives from orthographic variation. For example, the diacritized Sanadset form \textit{Mu\d{h}ammad ibn Ism\={a}'\={i}l} (with full tashkeel and Alef-with-hamza-below) and the undiacritized hawramani form \textit{Mu\d{h}ammad ibn Ismail} (with bare Alef) map to identical strings under $\mathrm{norm}(\cdot)$.

% ── 4.2 Candidate Generation ──────────────────────────────────────────────────
\subsection{Candidate Generation}
\label{sec:candidates}

A direct O($n \times m$) comparison of all 185,216 Sanadset narrators against all hawramani or muslimscholars records is computationally prohibitive. Both linking phases use a \textbf{bigram prefix index}: each database is indexed by pairs of consecutive normalized Arabic tokens (bigrams of the name). For a given query name, candidate records are retrieved as the union of all indexed records sharing at least one bigram with the query. Typical candidate set size is a few dozen to a few hundred records per query, reducing the comparison burden by three to four orders of magnitude.

For Phase~2, common names such as \textit{Mu\d{h}ammad} appear in approximately 920 muslimscholars records. When hawramani provides a death year, candidate muslimscholars records are pre-filtered to those within $\pm$1 decade; when no death year is available, all candidates sharing the first name token are considered.

% ── 4.3 Phase 1: Sanadset → Hawramani ────────────────────────────────────────
\subsection{Phase 1: Sanadset~$\to$~Hawramani (Name-Only)}
\label{sec:phase1}

Sanadset narrator entries carry no biographical metadata---no death year, no grade, no source citations. Matching is therefore restricted to name similarity alone. For each Sanadset narrator $q$ and each hawramani candidate $c$ retrieved by bigram indexing, similarity is computed using token-sorted fuzzy string matching \citep{christen2012data}: tokens are sorted alphabetically before edit-distance comparison, making the score insensitive to word-order variation in Arabic nasab chains. Denoting this score as $\mathrm{TSR}(\cdot,\cdot) \in [0,1]$:

\begin{equation}
  s_{\text{name}}(q, c) = \mathrm{TSR}\!\bigl(\mathrm{norm}(q),\; \mathrm{norm}(c)\bigr)
\end{equation}

The best-scoring candidate above $s_{\text{name}} \geq 0.80$ is assigned as the link. Confidence tiers for Phase~1 are:

\begin{equation}
  \text{conf}_1(s) =
  \begin{cases}
    \HIGH  & s \geq 0.90 \\ \MED   & s \geq 0.80 \\ \textbf{NONE} & s < 0.80
  \end{cases}
\end{equation}

The threshold of 0.80 was chosen by inspection to balance recall (capturing partial name matches due to nasab chain truncation in Sanadset) against precision (rejecting near-homonyms); formal calibration against a manually annotated evaluation set is deferred to future work (Section~\ref{sec:discussion}). Narrators with no bigram candidate in hawramani (88,343 of 185,216; 47.7\%) receive no attempt at matching.

% ── 4.4 Phase 2: Hawramani ↔ Muslimscholars ──────────────────────────────────
\subsection{Phase 2: Hawramani~$\leftrightarrow$~Muslimscholars (Multi-Signal)}
\label{sec:phase2}

The hawramani--muslimscholars cross-reference uses three signals weighted according to data availability.

The three signals are combined into a weighted score. Because hawramani death years are available for fewer than 0.1\% of matched pairs, the weight on name similarity is effectively 0.70 for 99.9\% of the corpus---the rare case where both sides carry a death year reduces the name weight to 0.50 while increasing the death-year weight to 0.40. Death-year proximity
\label{sec:scoring-death}
is the strongest disambiguator for common Arabic names when available; muslimscholars provides death years for approximately 21.7\% of entries in matched pairs. The proximity score is:

\begin{equation}
  s_{\text{death}}(\Delta y) =
  \begin{cases}
    1.00 & \Delta y = 0 \\ 0.85 & \Delta y \leq 2 \\ 0.60 & \Delta y \leq 5 \\ 0.30 & \Delta y \leq 10 \\ 0.00 & \Delta y > 10
  \end{cases}
\end{equation}

where $\Delta y = |y_{\text{HW}} - y_{\text{MS}}|$ in Hijri years. Historical Hijri dates have transcription variance of $\pm$1--2 years across classical sources, justifying the 0.85 score at $\Delta y \leq 2$. When either side lacks a death year, $s_{\text{death}}$ takes the neutral value 0.50. Narrator reliability grades (weight 0.10) from hawramani and muslimscholars are collapsed to three polarity classes---\textit{positive}, \textit{negative}, and \textit{unknown}---and scored 1.0, 0.0, and 0.5 respectively for consistent, contradictory, and unknown polarity.

Because hawramani death years are sparse, the combined scoring function uses dynamic weights:

\begin{equation}
  \text{score} =
  \begin{cases}
    0.50 \cdot s_{\text{name}} + 0.40 \cdot s_{\text{death}} + 0.10 \cdot s_{\text{grade}} + \delta_{\text{src}} & \text{both sides have death year} \\[4pt] 0.70 \cdot s_{\text{name}} + 0.20 \cdot s_{\text{death}} + 0.10 \cdot s_{\text{grade}} + \delta_{\text{src}} & \text{otherwise}
  \end{cases}
  \label{eq:score}
\end{equation}

where $\delta_{\text{src}} = 0.05$ if the hawramani record has five or more source citations, else 0 (five citations is an approximate median of source counts in the hawramani database, distinguishing well-attested records from sparsely-documented ones). Because 99.9\% of pairs fall into the ``no year'' branch, the effective weight distribution is approximately: name~70\%, death~10\% (neutral $0.50 \times 0.20$), grade~10\%, source bonus up to~5\%. Confidence tiers follow $\text{score}$ thresholds: \HIGH{} $\geq 0.85$; \MED{} $\geq 0.65$; \LOW{} $\geq 0.45$; \textbf{NONE} below 0.45. These boundaries correspond to natural breaks in the observed score distribution and were set by inspection; formal precision--recall calibration is deferred to future work (Section~\ref{sec:discussion}).

% ── 4.5 Chain Composition and Graph Construction ──────────────────────────────
\subsection{Transmission Graph Construction and Enrichment}
\label{sec:graph}

The directed transmission graph is built by extracting consecutive narrator pairs within each sanad chain: if narrator~$A$ immediately precedes narrator~$B$, a directed edge $A \to B$ (student~$\to$~teacher) is added or its weight incremented, yielding 814,093 unique directed edges among 185,216 nodes with edge weights representing co-occurrence frequency across the corpus.

Node metadata is assembled from the narrator index---bare normalized name, English and Indonesian transliterations, total Hadith count, and book count---and extended with hawramani URL, Arabic name, death year, and reliability grades, and muslimscholars ID and English name, for all nodes with a confirmed Phase~1 link. The enriched graph is serialized as Parquet for sub-second loading after an initial build of approximately 30~seconds.

For Phase~1-linked narrators whose hawramani record is itself linked in Phase~2, a \textit{chain confidence} is assigned as the minimum of the two constituent confidence tiers (where HIGH $>$ MED $>$ LOW), so that the weaker leg determines the overall chain reliability.

% ============================================================
\section{Results}
\label{sec:results}

\subsection{Phase 1: Sanadset~$\to$~Hawramani Link Corpus}
\label{sec:results-p1}

\begin{table}[H]
\centering
\caption{Phase~1 (Sanadset~$\to$~hawramani) linking results.}
\label{tab:p1}
\begin{tabular}{lrr}
\toprule
Metric & Count & \% of 185,216 \\
\midrule
Total Sanadset unique narrators & 185,216 & 100.0\% \\ \quad With bigram candidates & 96,873 & 52.3\% \\ \quad\quad Linked (HIGH) & 39,938 & 21.6\% \\ \quad\quad Linked (MEDIUM) & 54,690 & 29.5\% \\ \quad\quad Below threshold (NONE) & 2,245 & 1.2\% \\ \quad No bigram candidates & 88,343 & 47.7\% \\
\midrule
\textbf{Total linked} & \textbf{94,628} & \textbf{51.1\%} \\
\bottomrule
\end{tabular}
\end{table}

Of 185,216 Sanadset narrator name variants, 94,628 (51.1\%) are linked to a hawramani record, with 39,938 at HIGH confidence ($s_{\text{name}} \geq 0.90$) and 54,690 at MEDIUM ($s_{\text{name}} \geq 0.80$). The 88,343 narrators with no bigram candidates represent names without any two-token overlap with any hawramani entry---typically highly abbreviated forms (single tokens like \textit{abiyhi} (``his father'')), kunyah-only forms with no ism present, or narrators genuinely absent from hawramani.

\subsection{Phase 2: Hawramani~$\leftrightarrow$~Muslimscholars Link Corpus}
\label{sec:results-p2}

\begin{table}[H]
\centering
\caption{Phase~2 (hawramani~$\leftrightarrow$~muslimscholars) linking results.}
\label{tab:p2}
\begin{tabular}{lrr}
\toprule
Metric & Count & \% of 100,915 \\
\midrule
Total hawramani narrators & 100,915 & 100.0\% \\ \quad Linked (HIGH, score $\geq 0.85$) & 18,245 & 18.1\% \\ \quad Linked (MEDIUM, score $\geq 0.65$) & 71,546 & 70.9\% \\ \quad Linked (LOW, score $\geq 0.45$) & 5,782 & 5.7\% \\ \quad No MS candidates & 5,113 & 5.1\% \\ \quad Below threshold & 229 & 0.2\% \\
\midrule
\textbf{Total linked} & \textbf{95,573} & \textbf{94.7\%} \\
\bottomrule
\end{tabular}
\end{table}

The very high coverage (94.7\%) reflects that muslimscholars, with 25,247 entries, is a curated subset of well-documented narrators who are virtually all present in hawramani (100,915 entries) by construction. The confidence distribution skews toward MEDIUM (70.9\%): for the 71.6\% of matched pairs lacking a hawramani death year, the effective scoring function is name-dominated (Eq.~\ref{eq:score}), producing many scores in the 0.65--0.85 range. The 18.1\% HIGH tier reflects pairs where death year data (available for 27.2\% of hawramani narrators) provided additional discrimination. The 5,113 hawramani narrators with no muslimscholars candidate tend to be obscure local transmitters or those documented only in rare sources not covered by muslimscholars. The full Phase~2 run completed in 165.2~seconds on a single CPU core.

\begin{figure}[!ht]
\centering
\includegraphics[width=0.9\textwidth]{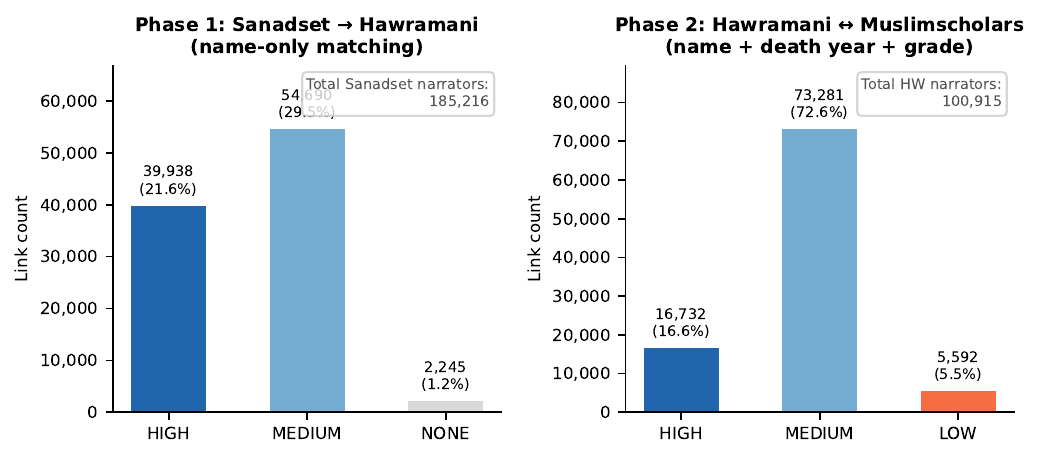}
\caption{Confidence tier distributions for Phase~1 (left) and Phase~2 (right).}
\label{fig:confidence}
\end{figure}

Figure~\ref{fig:confidence} shows the confidence tier distributions for both phases. The left panel gives Phase~1 counts as percentages of 185,216 total Sanadset narrators; the right panel gives Phase~2 counts as percentages of 100,915 hawramani entries. MEDIUM dominates Phase~2 (70.9\%), though HIGH has grown to 18.1\% as improved death year extraction now covers 27.2\% of hawramani narrators. For the 71.6\% of matched pairs lacking a hawramani death year, scoring collapses to the name-dominated regime (Eq.~\ref{eq:score}).

\subsection{Transmission Graph}
\label{sec:results-graph}

\begin{figure}[!ht]
\centering
\includegraphics[width=\textwidth]{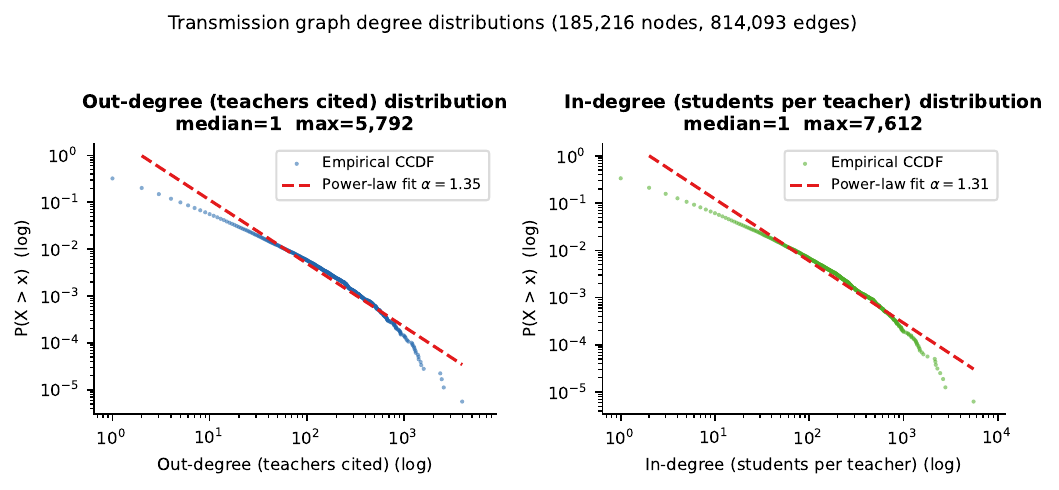}
\caption{Complementary cumulative degree distributions (CCDF) of the transmission
graph on log-log axes; dashed lines show power-law fits.}
\label{fig:degree}
\end{figure}

The directed transmission graph of 185,216 nodes and 814,093 edges exhibits structural properties consistent with prior, smaller-scale analyses of Hadith transmission networks \citep{saeed2021sna,alam2021sna} (Table~\ref{tab:graph}). Both in- and out-degree distributions approximate power laws with heavy tails (Figure~\ref{fig:degree}), consistent with scale-free network properties. The graph is approximately 90 times larger in node count than the next largest published Hadith narrator graph \citep{multiisnadset2024} (2,092 nodes) and spans 926 books compared to single-collection analyses. The highest in-degree narrators include Ab\={u} Hurayra (2,789) and Ibn `Abb\={a}s (2,636). The two entries with the highest in-degree overall---\textit{abiyhi} (``his father''; 7,612) and \textit{ab\={\i}} (``my father''; 5,550)---are relational tokens, not proper narrator names; they aggregate edges that refer to many distinct individuals. Downstream centrality analyses should filter these nodes before interpreting degree-based rankings.

\begin{table}[!ht]
\centering
\caption{Transmission graph statistics.}
\label{tab:graph}
\begin{tabular}{lr}
\toprule
Metric & Value \\
\midrule
Nodes (unique normalized narrator names) & 185,216 \\ Directed edges (student~$\to$~teacher pairs) & 814,093 \\ Nodes with at least one out-edge & 178,511 (96.4\%) \\ Nodes with at least one in-edge & 158,920 (85.8\%) \\ Out-degree: median / mean / max & 1 / 4.6 / 5,792 \\ In-degree: median / mean / max & 1 / 5.1 / 7,612 \\
\bottomrule
\end{tabular}
\end{table}

Of the 94,628 Phase~1-linked narrators, 93,588 (98.9\%) have a hawramani record that is also linked in Phase~2, yielding 93,588 three-source alignments (Sanadset~$\to$~hawramani~$\to$~muslimscholars): \HIGH{} 12,911 / \MED{} 78,841 / \LOW{} 1,836, stratified by chain confidence (minimum of the two constituent tiers; see Section~\ref{sec:graph}). The remaining 1,040 Phase~1-linked narrators have no muslimscholars counterpart in Phase~2.

\begin{figure}[t]
\centering
\includegraphics[width=\textwidth]{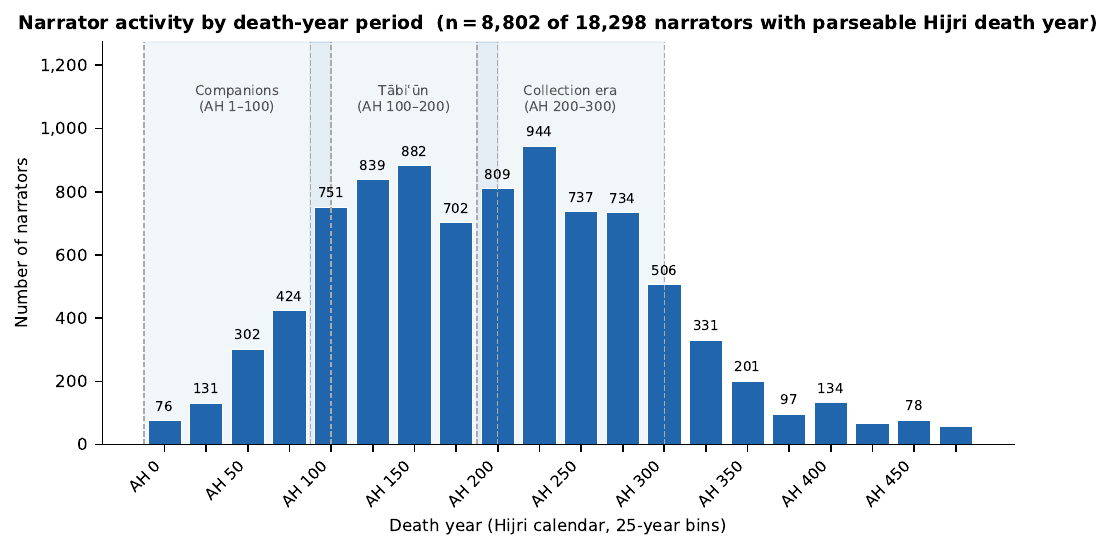}
\caption{Narrator activity by death-year period (AR-Sanad narrators\_db,
$n = 8{,}802$ narrators with parseable Hijri death year, 25-year bins).}
\label{fig:timeline}
\end{figure}

Using the narrators\_db from AR-Sanad \citep{mahmoud2022arsanad}, which provides biographical records for 18,298 narrators, Figure~\ref{fig:timeline} shows the distribution of narrator activity across 25-year Hijri bins for the 8,802 narrators with a parseable Hijri death year (AH~1--500). Activity peaks across the AH~100--225 range (five consecutive 25-year bins), corresponding to the generations of the \textit{T\={a}bi`\={u}n} (Successors) and \textit{Atb\={a}` al-T\={a}bi`\={i}n} (Successors of Successors)---the periods of most intense Hadith transmission and collection. The Companions' generation (AH~1--100) shows moderate representation, reflecting their smaller number relative to later generations.

% ============================================================
\section{Discussion}
\label{sec:discussion}

\subsection{Challenges and Comparison to General Arabic Entity Resolution}

The entity resolution task for Hadith narrators is qualitatively harder than general Arabic named entity disambiguation \citep{elzanfaly2016arabic} for reasons that domain-general tools do not address. A narrator known in Sanadset as \textit{Mu\d{h}ammad ibn Ismail} may appear in hawramani as \textit{Mu\d{h}ammad ibn Ismail ibn Ibrahim ibn al-Mugh\={\i}ra al-Ju`f\={\i} al-Bukh\=ar\={\i}}. The additional nasab tokens dilute string similarity; token-sorted fuzzy matching partially mitigates token-order variation but the asymmetry in chain depth remains a systematic source of false negatives affecting the 88,343 Sanadset narrators with no bigram candidates (47.7\%). Kunyah/ism alternation compounds the problem: a narrator listed under kunyah (\textit{Ab\=u \d{H}an\=ifa}) has no token overlap with the same narrator's hawramani entry under ism (\textit{al-Nu\`man ibn Thabit}), and string-based methods assign such pairs a score near zero. The MEDIUM confidence dominance in Phase~2 (70.9\%) is not a reflection of intrinsic ambiguity but of data sparsity: only 4.8\% of matched pairs have death years on both sides (Section~\ref{sec:phase2}), and for the 71.6\% lacking a hawramani death year the scoring function collapses to the name-dominated branch of Eq.~\ref{eq:score}.

Domain-general Arabic NED \citep{elzanfaly2016arabic} relies on entity descriptions from Wikipedia-aligned knowledge bases unavailable for historical Islamic narrators. The pipeline presented here operates entirely on structured name strings and categorical metadata, making it applicable to any pair of Arabic biographical databases. The AR-Sanad v2 approach \citep{mahmoud2024arsanadv2} improves disambiguation within a closed narrator class set; our work addresses the orthogonal open-world linking problem, and the \sanadlinks{} and \narratorlinks{} corpora could serve as training data for a future neural entity linker.

\subsection{Limitations and Future Work}

Confidence thresholds were set by design and inspection; no formally annotated ground-truth evaluation set exists for Hadith cross-source narrator links. A set of 1{,}000--2{,}000 manually verified links evaluated by \textit{`ilm al-rij\={a}l} scholars is a priority for future work and would enable rigorous precision-recall calibration. A kunyah-to-ism mapping layer using narrators\_db \textit{shuhra} fields is the most direct path to recovering a substantial fraction of the 88,343 currently unlinked Sanadset narrators.

Death year sparsity in hawramani substantially weakens Phase~2 disambiguation for common names. Cross-referencing hawramani with narrators\_db death years \citep{mahmoud2022arsanad} or with OpenITI corpus extractions \citep{romanov2017algorithmic} could impute missing values and restore the temporal signal's discriminative power. The transmission graph also encodes structural information---shared teacher/student neighbours---that could supplement name-based matching via constraint propagation: if narrator~$A$ is HIGH-confidence matched and teacher~$B$ appears in hawramani as $A$'s known teacher, then $B$'s candidate match could be constrained to $A$'s documented teachers in the Phase~2 graph. This graph-structural approach is deferred to subsequent work.

% ============================================================
\section{Conclusion}
\label{sec:conclusion}

We presented a two-phase entity resolution pipeline that produces (a)~94,628 cross-source links between the Sanadset 650K corpus and the hawramani biographical database (51.1\% of 185,216 narrator name variants), and (b)~95,573 links between hawramani and the muslimscholars database (94.7\% of 100,915 hawramani entries). The pipeline combines domain-specific Arabic name normalization with a name-only matching phase (for Sanadset narrators, which carry no biographical metadata) and a multi-signal weighted scoring phase (for hawramani--muslimscholars pairs, combining name similarity, death-year proximity, and reliability grade polarity). Both corpora are stratified into confidence tiers; the \HIGH-confidence subset of the Sanadset--hawramani links (39,938 narrators, $s_{\text{name}} \geq 0.90$) represents the most reliable starting point for downstream knowledge graph integration.

The linked corpora enable construction and enrichment of a 185,216-node, 814,093-edge directed transmission graph---the largest published Hadith narrator network by node count---with cross-source biographical metadata enabling analyses not achievable from Sanadset data alone.

The contributions address a gap identified by multiple prior surveys: no cross-source narrator entity resolution resource existed at meaningful scale. The \sanadlinks{} corpus, \narratorlinks{} corpus, and enriched narrator graph are released as open datasets to support downstream research in Hadith authentication, digital Islamic humanities, and Arabic historical NLP.

% ============================================================
\section*{Data Availability}

The \sanadlinks{} and \narratorlinks{} link corpora and enriched narrator graph are released as CSV files at \citet{wirahman2026dataset} (\url{https://doi.org/10.5281/zenodo.21019693}). The Sanadset 650K corpus is available from its original publication \citep{mghari2022sanadset}. The AR-Sanad narrators\_db is distributed with \citep{mahmoud2022arsanad}. Hawramani data were scraped in June 2026; the underlying biographical texts are sourced from the Shamela digitisation project~\citep{hawramani2026} and comprise classical Islamic works predating the 20th century, which are in the public domain. Muslimscholars data were scraped in June 2026; users requiring redistribution of that content should verify terms with the platform operators~\citep{muslimscholars2026}.

% ============================================================
\bibliographystyle{plainnat}
\bibliography{refs}

\end{document}